\documentclass[
                reqno,         
               finalverso,       
]{nrc2}                          

\usepackage{graphicx,color}                  

\usepackage{amsmath}  
\usepackage{amssymb}        
\usepackage{bm}             

\usepackage{url}            
\NRCurl{phys}

\usepackage[space]{cite}           

\usepackage[french,english]{babel}

    \large

\renewcommand{\eqnoformat}[1]{(#1)}

\def\be{\begin{equation}}
\def\ee{\end{equation}}
\def\ba{\begin{eqnarray}}
\def\ea{\end{eqnarray}}

\begin{document}




\title{Primordial Magnetism in CMB B-modes}

\author[Levon Pogosian]{Levon Pogosian}
\address{Department of Physics, Simon Fraser University, Burnaby, BC, V5A 1S6, Canada, and
DAMTP, University of Cambridge, Cambridge, CB3 0WA, UK.}
\correspond{levon@sfu.ca}

\author[Tanmay Vachaspati]{Tanmay Vachaspati}
\address{Physics Department, Arizona State University, Tempe, AZ 85287, USA.}

\author[Amit Yadav]{Amit Yadav}
\address{Institute for Advanced Study, Princeton, NJ 08540, USA, and Center for Astrophysics and Space Sciences, University of California, San Diego, 9500 Gilman Drive, La Jolla, CA, 92093-0424.}

\shortauthor{L. Pogosian, T. Vachaspati, and A.~Yadav}


\begin{abstract}
Cosmic Microwave Background (CMB) polarization B-modes induced by Faraday Rotation (FR) can provide a distinctive signature of primordial magnetic fields because of their characteristic frequency dependence and because they are only weakly damped on small scales. FR also leads to mode-coupling correlations between the E and B type polarization, and between the temperature and the B-mode. These additional correlations can further help distinguish magnetic fields from other sources of B-modes. We review the FR induced CMB signatures and present the constraints on primordial magnetism that can be expected from upcoming CMB experiments. Our results suggest that FR of CMB will be a promising probe of primordial magnetic fields.
\end{abstract}
%

\maketitle


\section{Introduction}

Most of the ongoing theoretical and observational effort in cosmology is driven by its three most famous mysteries: Inflation, Dark Matter and Dark Energy. In comparison, the unexplained origin of large scale cosmic magnetic fields may seem like many other problems in astrophysics -- technically difficult but of lesser fundamental interest. This may, to some extent, be the case for magnetic fields in mature galaxies, where they could be generated through a dynamo mechanism~\cite{Widrow:2002ud}. However, explaining their presence in young protogalaxies, clusters and, possibly, in the intergalactic space \cite{Neronov:1900zz}, is more challenging. Evolution of magnetism in forming cosmic structures is highly non-linear which makes it difficult to conclusively rule out the existence of yet unknown astrophysical mechanisms that could generate magnetic fields on large scales and high redshifts. Another possibility is that magnetic fields existed before astrophysical structures began to form~\cite{Grasso:2000wj}. A primordial magnetic field (PMF) could be produced in the aftermath of cosmic phase transitions \cite{Vachaspati:1991nm,Cornwall:1997ms} or in specially designed inflationary scenarios \cite{Turner:1987bw,Ratra:1991bn}. Measurements of cosmic microwave background (CMB) temperature and polarization could decisively prove their primordial origin if they contained magnetic signatures present at the time of last scattering. A discovery of PMF would have profound implications for our understanding of the early universe, with critical insights into fundamental problems such as the matter-antimatter asymmetry~\cite{Vachaspati:2001nb}.

A stochastic PMF influences CMB observables in several ways. Magnetic stress-energy perturbs the metric which leads to CMB anisotropies, while the Lorentz force deflects moving electrons and protons coupled to photons. Recently, it has been suggested \cite{Jedamzik:2011cu} that small-scale fields can appreciably alter the recombination history and, consequently, the distance to last scattering, because of the enhanced small scale baryonic inhomogeneities. Here we focus on another signature of PMF -- the Faraday Rotation (FR) of CMB polarization.

FR produces a $B$ mode type polarization with a characteristic 
spectrum \cite{Kosowsky:2004zh,Pogosian:2011qv} as well as non-trivial 4-point correlations of the CMB temperature and polarization. In \cite{Yadav:2012uz}, we examined detectability of PMF using different correlators and evaluated their relative merits. We fond that a Planck-like experiment can detect scale-invariant PMF of 
nG strength using the FR diagnostic at $90$GHz, while realistic future experiments at the same frequency can detect $10^{-10}\, {\rm G}$. This is comparable or better than other CMB probes of PMF, and using multiple frequency channels can further improve on these prospects.

\section{B-modes from Faraday Rotation}

At a given direction $\hat{\bf n}$ on the sky, CMB is characterized by its intensity and two additional Stokes parameters, $Q$ and $U$, quantifying its linear polarization. While $Q(\hat{\bf n})$ and $U(\hat{\bf n})$ are the quantities that experiments directly measure, their values depend on the choice of the coordinate axes. Instead, it has become customary to interpret polarization maps by separating them into parity-even and parity-odd patterns, or the so-called $E$ and $B$ modes~\cite{Kamionkowski:1996zd,Seljak:1996gy}. Existence of intensity fluctuations at last scattering implies generation of $E$ modes, which by now have been observed and found to be consistent with the spectrum of temperature anisotropies. On the other hand, $B$ modes would not be generated at last scattering unless there were gravitational waves or other sources of metric perturbations with parity-odd components such as cosmic defects \cite{Seljak:1997ii} or magnetic fields \cite{Seshadri:2000ky}. Weak lensing (WL) of CMB photons by the large scale structures along the line of sight distorts polarization patterns generated at last scattering and converts some of the $E$ mode into $B$ modes, which is expected to be measured with upcoming CMB experiments.

A primordial magnetic field present at and just after last scattering will Faraday-rotate the plane of polarization of the CMB photons. The rotation angle along $\hat{\bf n}$ is given by
\begin{equation}
\alpha(\hat{\bf n}) = \frac{3}{{16 \pi^2 e}} \lambda_0^2 
\int \dot{\tau} \ {\bf B} \cdot d{\bf l} \ ,
\label{theta2}
\end{equation}
where $\dot{\tau} \equiv n_e \sigma_T a$ is the differential optical depth, $n_e$ is the line of sight free electron density, $\sigma_T$ is the Thomson scattering cross-section, $a$ is the scale factor, $\lambda_0$ is the observed wavelength of the radiation, ${\bf B}$ is the ``comoving'' magnetic field, and $d{\bf l}$ is the comoving length element along the photon trajectory. 

Statistically homogeneous, isotropic and Gaussian distributed stochastic magnetic fields can be characterized by a two-point correlation function in Fourier space 
\be
\langle b_i ({\bf k} ) b_j ({\bf k}' ) \rangle =
(2\pi )^3 \delta^{(3)}({\bf k} + {\bf k}' )
[ (\delta_{ij} - {\hat k}_i {\hat k}_j) S(k) 
\label{bcorr}
\ee
where $S(k)$ is the symmetric magnetic power spectrum, and where we omit the anti-symmetric contribution that quantifies the amount of magnetic helicity because only $S(k)$ contributes to the CMB observables evaluated in this paper. The shape of $S(k)$ depends on the mechanism responsible for production of PMF and generally can be taken to be a power law up to a certain dissipation scale:
\be
S(k) 
\propto \begin{cases}
k^{2n-3} &\mbox{$0<k<k_{\rm diss}$}\\ 
 0 &\mbox{$k>k_{\rm diss}$}
 \end{cases} \ .
\label{eq:singlePB}
\ee 
The dissipation scale, $k_{\rm diss}$, should, in principle, be dependent on the amplitude and the shape of the magnetic fields spectrum. We assume that $k_{\rm diss}$ is determined by damping into Alfven waves \cite{Jedamzik:1996wp,Subramanian:1997gi} and can be related to $B_{\rm eff}$ as
\be
{k_{\rm diss} \over 1{\rm Mpc}^{-1}} \approx
          1.4 \ h^{1/2} \left( 10^{-7} {\rm Gauss} \over B_{\rm eff} \right) \ ,
\label{kIBeff}
\ee
where $B_{\rm eff}$ is defined as the effective homogeneous field strength that would have the same total magnetic energy density. $B_{\rm eff}$ is related to the fraction of magnetic energy density to the total radiation density, $\Omega_{B \gamma}$, via \cite{Pogosian:2011qv}
\be
B_{\rm eff}= 3.25 \times 10^{-6} \sqrt{\Omega_{B \gamma}} \ {\rm Gauss} \ .
\label{beff-omega}
\ee

The generation of CMB polarization and the FR happen concurrently during the epoch of last scattering. However, as we have shown in \cite{Pogosian:2011qv}, assuming an instantaneous last scattering, i.e. that $E$ modes were produce first and subsequently rotated by PMF, results in relatively minor inaccuracies. In this approximation, the relation between the spherical expansion coefficients of the $E$, $B$ and $\alpha$ fields can be written as
\be
B_{lm}= \sum_{LM}\sum_{l_1 m_1}\alpha_{LM} E_{l_1 m_1} {\cal M}^{LM}_{l_1m_1} \ ,
\label{eq:blm}
\ee
where ${\cal M}^{LM}_{l_1m_1}$ is defined in terms Wigner $3$-$j$ symbols~\cite{Kamionkowski:2008fp}. We note that $B$ modes from WL can also be schematically written as (\ref{eq:blm}) but with a different mixing matrix 
${\cal M}^{LM}_{l_1m_1}$. Importantly, the mixing matrix for WL has a parity opposite to that of FR so that the two rotations are orthogonal to each other, making it possible to reconstruct them separately.

\begin{figure}[tbp]
\includegraphics[height=0.48\textwidth]{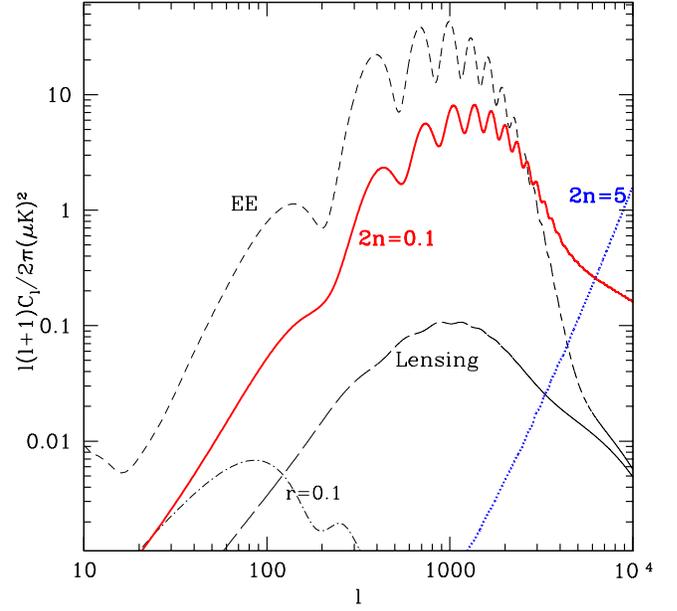}
\caption{The CMB B-mode spectrum from Faraday rotation in the case of a nearly scale-invariant magnetic spectrum with $2n=0.1$ and $\Omega_{B\gamma}=5\times 10^{-4}$ (solid red), and a causal spectrum with $2n=5$ and 
$\Omega_{B\gamma}=10^{-3}$, at $61$~GHz. The black short-dash line is the input E-mode spectrum, the black dash-dot line is the contribution from inflationary gravitational waves with $r=0.1$, while the black long-dash line is the expected contribution from gravitational lensing by large scale structure.}
\label{fig:cl}
\end{figure}

In Fig.~\ref{fig:cl} we show the B-mode auto-correlation spectra due to FR by stochastic magnetic fields with two different spectra. One, with $2n \approx 0$, corresponds to nearly scale-invariant PMF generated via an inflationary mechanism  \cite{Turner:1987bw,Ratra:1991bn}, while the other, with $2n=5$, corresponds to PMF produced causally in phase transitions \cite{Durrer:2003ja,Jedamzik:2010cy}, e.g. at the time of electroweak or QCD symmetry breaking. Also shown is the $E$ mode auto-correlation spectrum which acts as a source for the FR $B$ modes, as well as $B$ modes from inflationary gravitational waves with $r=0.1$, and the expected contribution from WL.

The FR induced $B$ mode spectra have certain characteristic features. In the case of the nearly scale 
invariant magnetic spectrum, the spectrum is oscillatory. In fact, the shape of the B-mode spectrum mimics that of the E-mode, except for the lack of exponential damping on small scales. This is because $E$ modes are suppressed by the Silk damping, while PMF can remain correlated on small scales. The damping of the FR induced B-mode power is due to averaging over many random rotations along the line of sight. This translates into a $1/l$ suppression of the angular spectrum, i.e. asymptotically we have $l^2 C_l^{BB} \propto l^{2n-1}$ at large $l$. In the case of $2n=5$, expected for causally generated magnetic fields~\cite{Jedamzik:2010cy}, the FR produced B-mode can dominate the signal at high $l$. It is thus interesting to consider possibility of future $B$ mode experiments specially designed to look for cosmological signals at sub-arcmin scales.

\section{Signal in mode-coupling correlations}

\begin{figure}[tbp]
\includegraphics[height=0.5\textwidth]{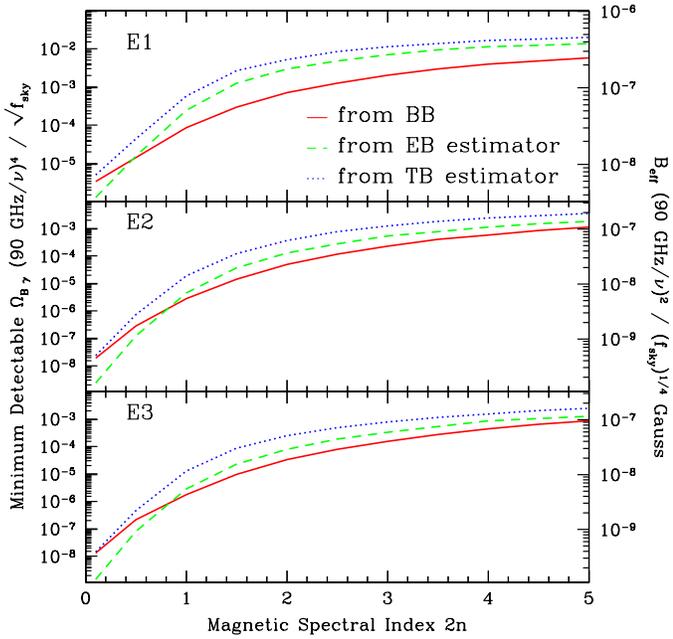}
\caption{The minimum detectable magnetic field amplitude $\tilde{\Omega}_{B\gamma}$ as a function of the magnetic spectral index $2n$ for the three estimators, $BB$ (solid red), $EB$ (dashed green) and $TB$ (dotted blue). The top panel is for E1, the middle panel is for E2 and the lower panel is for E3.}
\label{fig:ovsn}
\end{figure}

Spatially dependent FR produces additional non-Gaussian signatures in the CMB polarization. Namely, a particular realization of the FR distortion field correlates the respective Legendre coefficients $E_{lm}$ and $B_{l'm'}$. In fact, as shown in \cite{Kamionkowski:2008fp}, it is possible to reconstruct the distortion field $\alpha(\hat{\bf n})$ from specially constructed linear combinations of products $E_{lm}B_{l'm'}$. The additional correlations induced by FR also manifest themselves as connected 4-point functions of the CMB, which, in turn, provide a measurement of the distortion spectrum $C_L^{\alpha \alpha}$ \cite{Yadav:2009eb,Gluscevic:2009mm}. In principle, one can construct four estimators of the distortion spectrum, based on products of two CMB fields one of which contains polarization: $TE, EE, TB,$ and $EB$. Of these four, the first two receive a large contribution to their variance from the usual scalar adiabatic Gaussian perturbations which makes it harder to find the FR signal. In \cite{Yadav:2012uz}, we studied the last two, i.e we considered estimators based on 4-point correlations $\langle EBEB \rangle$ and $\langle TBTB \rangle$, and compared them to the 2-point function $\langle BB \rangle$. We asked which of the estimators has the highest signal to noise for several types of magnetic field spectra and for a range of experimental sensitivities. We have fully accounted for the contamination by weak lensing, which contributes to the variance, but whose contribution to the 4-point correlations is orthogonal to that of FR. 

To forecast the detectability of FR we looked at three experimental setups: a Planck-like satellite \cite{Planck:2006aa} (E1), a ground- or balloon-based experiment realistically achievable in the next decade (E2), and a future dedicated CMB polarization satellite (E3).  The forecasts depend on the fraction of the sky covered by the experiment, $f_{\rm sky}$,  which is close to unity for E1 and E3, and will be smaller for E2. We present our forecasted bounds on the PMF fraction in Fig.~\ref{fig:ovsn} subject to specifying $f_{\rm sky}$, which only appears under a quartic root in the bounds on $B_{\rm eff}$. 

As Fig.~\ref{fig:ovsn} demonstrates, FR will be a very promising diagnostic of PMF. In particular, future generation of sub-orbital or space-based CMB polarization experiments will be able to detect scale-invariant magnetic fields as weak as $10^{-10}$G based on the measurement at $90$ GHz frequency. Measurements at multiple frequencies can further significantly improve on these prospects.

The relative strengths of the three estimators, demonstrated in Fig.~\ref{fig:ovsn} can be understood as follows. Generally, the $EB$ and $TB$ estimators have a larger number of independent modes contributing to the signal than the $B$-mode spectrum. Thus, in principle, it is not surprising if they result in a higher signal to noise. However, whether that is the case depends on the experimental noise level, and the distribution of power in the given combination of CMB fields and in the magnetic field. For a scale-invariant PMF spectrum, the $B$-mode is essentially a copy of the $E$-mode, with most of the $B$-mode power being on scales where the $E$-modes are also most prominent. This results in a strong correlation between $E$ and $B$ for scale-invariant fields. In the case of the $TB$ correlation, the underlying $T$ and $E$ ($B$ is obtained by a scale-invariant rotation of $E$) fields peak on rather different scales. Namely, $T$ peaks at $\ell \sim 200$ while $E$ peaks at $\ell \sim 1000$. In other words, the intrinsic correlation between $T$ and $E$ is already suboptimal, translating into a lesser correlation between $T$ and $B$. Thus, for experiments with sufficiently low noise levels, such as E1, E2 and E3 considered in this paper, the $EB$ estimator performs better than $TB$ for scale-invariant fields. This would not necessarily remain true if polarization measurements had a significantly higher experimental noise.

For causally generated PMF with steeply rising (``blue'') spectra, as in the $2n=5$ case, the FR power is concentrated on very small scales, far away from the scales at which any of the unrotated CMB fields have significant power. This means that the $B$-modes in the observable range are obtained either by a rotation of $E$-modes far away from their peak power scale, or by a rotation of peak $E$-mode by a negligible angle. This means that $E$ and $B$ fields peak at very different scales, with their correlation being close to zero over the observable scales. In this case, we see that the $B$-mode spectrum, i.e. the $BB$ correlation, has the highest signal to noise.

\section{Discussion and Outlook}

When interpreting the forecasted bounds on the magnetic field energy fraction or the 
effective magnetic field strength in Fig.~\ref{fig:ovsn}, several points must be kept in mind. 
First, the constraints are obtained after setting the dissipation scale to be given by Eq.~(\ref{kIBeff}). 
For scale-invariant fields it makes no difference, as in this case most of the signal is on scales larger than the magnetic dissipation scale, and $k_{\rm diss}$ does not contribute to the normalization of the spectra when $2n \rightarrow 0$. In fact, for scale-invariant fields, the effective field $B_{\rm eff}$ defined via Eq.~(\ref{beff-omega}) 
is the same as the commonly used $B_\lambda$, which is the field smoothed on a given scale $\lambda$. 
Thus, our forecasts of the minimum detectable $B_{\rm eff}$ for scale-invariant fields can be directly 
compared to most other bounds in the literature. 

CMB is less sensitive to magnetic fields with blue spectra because most of the 
anisotropies are concentrated on very small scales. This is what Fig.~\ref{fig:ovsn} is showing too. 
Here we note that Fig.~\ref{fig:ovsn} assumes that the spectrum will 
keep rising at the same steep rate ($2n=5$) all the way to the dissipation scale. 
On the other hand, simulations \cite{Jedamzik:2010cy} suggest that the spectrum becomes 
less steep, with $2n'=3$ over some intermediate range $k_I <k< k_{\rm diss}$, implying a smaller 
net magnetic energy fraction $\Omega_{B\gamma}$. Big Bang Nucleosynthesis constrains this fraction to be less than $10$\% \cite{Kernan:1995bz,Cheng:1996yi,Grasso:1996kk,Kawasaki:2012va}. Note that the expected bound from Planck (E1) in Fig.~\ref{fig:ovsn} for $2n=5$ will be at least an order of magnitude stronger, while E2 and E3 will improve on the BBN bound by two orders of magnitude. Here it is worth keeping in mind that our bounds are on the magnetic field contribution at the time of last scattering. While it is expected that the fields remain effectively frozen-in between the time of nucleosynthesis and last scattering, with a relatively slow time evolution of the dissipation scale, this is still an approximation. In any case, it is interesting to know how good the resolution of future $B$ mode experiments can be, since the FR contribution from causally generated PMF appears to dominate over other cosmological sources on small scales.

In the case of scale-invariant fields, existing bounds on the magnetic field strength from WMAP are at a 
level of a few nG \cite{Paoletti:2010rx}. These bounds are based on 
the anisotropies induced by the metric fluctuations sourced by magnetic fields, and ignore the FR effect. 
In Refs.~\cite{Kahniashvili:2008hx,Pogosian:2011qv} the WMAP bound using FR was 
obtained at the $10^{-7}\, {\rm G}$ level. As one can see from Fig.~\ref{fig:ovsn}, 
Planck (E1) can almost match today's bounds for scale invariant ($n=0$) fields using the $EB$ 
estimator at only one frequency, while future probes, such as E2 and E3, can improve the bounds 
by an order of magnitude! This suggests that the mode coupling estimators of FR can be a very 
powerful direct probe of scale-invariant PMF.

Finally, as mentioned already, the bounds are based on using 
a single frequency band, while using several bands can further improve the constraints. While our estimates look quite promising, they are still preliminary and ignore the potentially devastating foreground effects. Prior to reaching our detectors, CMB must pass through the magnetic field of our own galaxy and will experience FR in which $B$ modes are produced. It remains to be shown to what extent the FR due to the galaxy can be subtracted from the cosmological FR signal.

\end{document}